# How Do Non-Ideal UAV Antennas Affect Air-to-Ground Communications?

M. G. Khoshkholgh*, Keivan Navaie**, Halim Yanikomeroglu†, V. C. M. Leung*, and Kang. G. Shin††
* Department of Electrical and Computer Engineering, the University of British Columbia
** School of Computing and Communications, Lancaster University
† Department of System and Computer Engineering, Carleton University
†† Department of Electrical Engineering and Computer Science, The University of Michigan

*Abstract*—Analysis of the performance of Unmanned Aerial Vehicle (UAV)-enabled communications systems often relies upon idealized antenna characteristic, where the side-lobe gain of UAVs' antenna is ignored. In practice, however, side-lobe cause inevitable interference to the ground users. We investigate the impact of UAVs' antenna side-lobe on the performance of UAV-enabled communication. Our analysis shows that even for a very small antenna's side-lobe gain, the ground receiver can experience substantial interference. We further show that a rather large exclusion zone is required to ensure a sufficient level of protection for the ground receiver. Nevertheless, in a multiple-antenna setting for the ground users, even when such a large exclusion zone was in place, UAVs' antenna side-lobe creates a high level of correlation among the interference signals received across receive antennas. Such a correlation limits the system ability to exploit channel diversity in a multiple-antenna setting for improving capacity. We then quantify the impact of UAVs' antenna side-lobes on the overall system performance by deriving the corresponding loss of the achieved capacity in various communications environments. We provide a new quantitative insight on the cost of adopting non-ideal UAV antenna on the overall capacity. Our analysis also shows that the capacity loss can be confined by careful selection of system parameters.

## I. INTRODUCTION

Using Unmanned Aerial Vehicles (UAVs) equipped with wireless transceivers acting as drone base-stations (BSs) is proposed to enhance the ground users' connectivity in 4G & 5G systems, and beyond [1–3]. UAVs' potential for enhancing capacity/coverage of LTE/LTE-A has also been demonstrated in cases of UAV-assisted communications and cellular-connected drones [4–6].

Adopting directional antennas at the UAVs enables direct communication with the intended ground UEs while limiting interferences to other users. Assuming an ideal directional antenna with a given beam-width and no or extremely small side-lobes, each UAV transmitter forms its own exclusive coverage zone. In this exclusive zone, the ground users are not interfered with by the UAVs whose coverage zones do not overlap. The notion of exclusion zone is adopted broadly to study the various aspects of UAV-enabled communications.

In a multiple UAV communication setting, the authors of [7] proposed a scheme for improving the spatial reuse of spectrum by selecting the proper size of the exclusion zones. Building upon the premise of exclusive coverage zone with no cross-zone interference, the authors of [8, 9] explored the optimal positioning of UAVs in order to maximize the coverage probability. They adjust the UAVs' altitude based on the system parameters and the type of communication environment, such as sub-urban, urban, dense-urban, and high-rise. The exclusive coverage zone is also used in [10] for offloading overloaded BSs in hot-spots. [11] also investigated the coexistence of UAV and device-to-device (D2D) communications in an exclusive coverage zone. Techniques for optimal partitioning of the ground plane into exclusive coverage zones was proposed in [12], with the objective of providing coverage to all the ground users. Exclusive coverage zones were considered in [13] to support low-latency ultra-reliable UAV-enabled connectivity.

In practice, however, as also noted in [14], UAVs' antennas are not ideal. In fact, by increasing the UAVs' altitude, the ground receiver is likely to receive many interfering signals transmitted through the antenna side-lobe of other UAVs [3]. This issue becomes more detrimental as for UAVs' hovering above a certain altitude the A2G propagation channel becomes dominantly Line-of-Sight (LOS) [5, 15]. This reduces the path-loss attenuation for the interfering signals, and thus may affect the ground receivers' performance and therefore generate a substantial mismatch between theoretical and actual results. To address this issue, we investigate the effect of antenna side-lobe on the performance of ground receivers.

We consider a UAV-enabled communications system with non-ideal antennas and analyze the performance of a ground receiver with multiple receive antennas located in the exclusion zone of a given UAV. To understand the impact of the interfering signals, we then find the number of detectable interfering signals from other UAVs, or *outside-UAVs*, at the ground receiver, i.e., interfering signals with a power level above the ground receiver's sensitivity. Adopting tools of stochastic geometry, we find that the ground receiver is most likely to receive a rather large number of detectable interfering signals from the UAVs located outside its exclusion zone. By estimating the probability of having non-zero detectable interfering signals at the ground receiver, we introduce the required size of exclusion zone so as to keep this probability below a given performance threshold, $\epsilon$.

Our numerical evaluation shows that for antennas with a main-lobe to side-lobe gain ratio of $G/g = 2500$ in a sub-urban environment, the required radius of the exclusion zone is around 50km given $\epsilon = 0.05$. In an extreme case of a small

$G/g = 12.5$, the required radius of the exclusion zone is raised to 300km. The size of the required exclusion zone is also affected by the communication environment. For instance, in a high-rise communication environment where the interference signals are received through a non-LOS (NLOS) dominant link, the required radius of the exclusion zone is reduced approximately by 4x compared to the sub-urban environment.

Interfering signals also affect the performance of multiple-antennas communications between the UAV and the ground user. We show that even with an exclusion zone designed to keep the probability of having more than zero detectable interference signals higher than $\epsilon$, the aggregated received interference signal across the received antennas of the ground user becomes highly correlated. Our investigation indicates that this correlation might be as high as 90%, simply due to the aggregated LOS interference induced by the signal leakage from the side-lobes of outside-UAVs. Such a correlation is detrimental to the independence of the received attending signals across the multiple-antennas in the ground user and severely affects its ability to harvest the otherwise existent channel diversity. To provide quantitative insights, we derive the expected capacity loss, and analyze the impact of various system parameters on such a capacity loss. Our analysis shows that unlike the terrestrial communications in which non-LOS propagation is often dominant, designing UAV communication systems without considering a realistic antenna pattern might lead to a substantial capacity loss in real settings.

## II. SYSTEM MODEL

We consider a ground user located at the origin receiving signals from a UAV, namely a *supporting UAV*, located exactly above the origin at altitude $\tilde{H} > 0$. Associated with this transmitter-receiver link, we also consider an exclusion zone, where the receiver is assumed to be located at its center. To protect the ground receiver against interferences, other UAVs hovering above the inclusion zone, namely *outside-UAVs*, are not allowed to transmit. Relevant to our analysis is the projection of the exclusion zone on the ground which is a circular disk with radius $Z > 0$. We denote the exclusion zone by a disk, $\mathcal{B}_O(Z)$. Further, the outside of the exclusion zone is denoted by annulus $\overline{\mathcal{B}}_O(Z)$.

We assume that the locations of the outside-UAVs in the 3-D space follow a Homogenous Poisson Point Process (HPPP). In this model, $\Phi = \{(X_i, H) \in \mathbb{R}^3, i = 1, 2, \ldots : \|X_i\| > Z\}$, where $X_i \in \overline{\mathcal{B}}_O(Z)$ is the location of UAV $i$ in the 2-D plane, and $H$ is its altitude. For brevity, we further assume that the UAVs are all at the same constant altitude. The density of UAVs is $\lambda$ units per km$^2$. All UAVs are equipped with a directional antenna with a circular radiation pattern and beam-width of $\omega$. We denote the main-lobe and side-lobe antenna gain for UAV $X_i$ by $X_i$, by $G$ and $g \neq 0$, respectively, where $g \ll G$. The vertical angel between the receiver and the UAV $X_i$ is also denoted by $\psi_i = \tan^{-1}(H/\|X_i\|)$. The receiver located at the origin is within the main-lobe of UAV $i$, if $\psi_i > \pi/2 - \omega/2$, or equivalently for $\|X_i\| < \frac{H}{\tan(\pi/2 - \omega/2)}$.

Therefore, to ensure that the receiver does not receive interference from the main-lobe of UAV $X_i$, $Z$ should be set to $Z = \frac{H}{\tan(\pi/2 - \omega/2)}$. Therefore, the exclusion zone of $X_i$ is not in the main-lobe of the out-side UAVs.

TABLE I
AIR-TO-GROUND CHANNEL PARAMETERS [8].

|        | High-Rise | Dense-Urban | Urban | Sub-Urban |
|--------|-----------|-------------|-------|-----------|
| $\phi$ | 27.23     | 12.08       | 9.61  | 4.88      |
| $\psi$ | 0.08      | 0.11        | 0.16  | 0.43      |

The channel between the UAVs and the users on the ground, referred to as the *A2G channel*, is modeled as a combination of a large-scale path-loss attenuation and a small-scale fading component [9, 15]. The A2G channel operates in LOS/NLOS mode [9], and the occurrence of LOS mode is shown to be dependent, among other things, on the drone's height, elevation angle, and the type of communication environment, e.g., dense urban or sparse rural. The probability that the channel between UAV $X_i$ and the receiver is an LOS-dominant channel is often considered as the distance-dependent probability [8, 9]:

$$p_L(\|X_i\|) = \left(1 + \phi e^{-\psi\left(\frac{180}{\pi}\arctan(\frac{H}{\|X_i\|}) - \phi\right)}\right)^{-1}, \quad (1)$$

where $\|X_i\|$ is the 2-D Euclidian distance between the ground user and the drone $X_i$, and $\phi$ and $\psi$ are the channel parameters representing the characteristics of the communication environment. As shown in (1), the probability of experiencing an LOS-dominant channel is increased by increasing $H$. Using (1), the path-loss attenuation is:

$$L(\|X_i\|) =$$
$$\begin{cases} L_L(\|X_i\|) = \frac{K_L}{(\sqrt{H^2 + \|X_i\|})^{\alpha_L}} & \sim p_L(\|X_i\|), \\ L_N(\|X_i\|) = \frac{K_N}{(\sqrt{H^2 + \|X_i\|})^{\alpha_N}} & \sim p_N(\|X_i\|), \end{cases} \quad (2)$$

where $\alpha_L$ ($\alpha_N$) is the LOS (NLOS) path-loss exponent, and $K_L$ ($K_N$) is the corresponding intercept constant, $\alpha_L \ll \alpha_N$. Note that increasing $H$ results in a higher attenuation as the signal needs to travel farther, experiencing a greater power loss, or equivalently a higher transmission power is required by the UAVs. Nevertheless, since $\alpha_L \ll \alpha_N$, a larger $H$ might be advantageous as it may make the LOS component dominant. In practice, $H$ should be carefully designed to balance the required transmit power on one hand, and the channel attenuation advantage on the other hand, see, e.g., [8, 9].

Each receiver has $R \geq 2$ antennas, indexed by $r$. Small-scale power fading between UAV $X_i$ and the $r$-th antenna at the receiver, $V_{X_i,r}$ is modeled by Nakagami fading:

$$V_{X_i,r} = \begin{cases} V_{X_i,r}^L = \Gamma(\overline{V}_L, \frac{1}{\overline{V}_L}) & \sim p_L(\|X_i\|) \\ V_{X_i,r}^N = \Gamma(\overline{V}_N, \frac{1}{\overline{V}_N}) & \sim p_N(\|X_i\|), \end{cases} \quad (3)$$

where $\Gamma(a, b)$ is Gamma distribution with parameters $a$ and $b$. For UAV $X_i$, we assume that $V_{X_i,r}$ are independent $\forall r$. Parameters $a$ and $b$ depend on the LOS/NLOS status of the communication channel between UAV $X_i$ and the receiver. We also note the fading is often more severe in NLOS channels, so it is reasonable to assume $\overline{V}_L > \overline{V}_N$.

## III. COUNT OF INTERFERING SIGNALS

Ideally, the side-lobe is NULL, i.e., $g = 0$, the receiver does not detect any signal from outside-UAVs. This simplifying assumption is widely adopted in the related literature, see, e.g., [7, 9]. In practice, however, $g \neq 0$, but the antenna is designed such that $g/G \ll 1$. Therefore, the signal transmitted by the outside-UAVs through the side-lobes may cause interference at the receiver. To investigate the impact of antenna side-lobes, we obtain the number of detectable interfering outside-UAVs signals at the receiver, or *signal count*, $\Sigma$. A similar terminology is used in the LTE systems, where the signal count to quantify the interference of neighboring cells [3].

We evaluate the signal count based on the aggregated received power across receive antennas,

$$\Sigma = \sum_{X_i \in \Phi} \mathbf{1}\left(\sum_r gPL(\|X_i\|, H)V_{X_i,r} \geq \gamma\right), \quad (4)$$

where $\gamma > 0$ is the receiver's sensitivity.

We now show that the signal count $\Sigma$ in (4) is a Poisson random variable. For this, we derive its Laplace transform:

$$\mathcal{L}_\Sigma(t) = \mathbb{E} e^{-t \sum_{X_i \in \Phi} \mathbf{1}(\sum_r gPL(\|X_i\|,H)V_{X_i,r} \geq \gamma)}$$

$$= \mathbb{E} \prod_{X_i \in \Phi} \left( e^{-t} \mathbb{P}\left\{ \sum_r V_{X_i,r} \geq \frac{\gamma}{gPL(\|X_i\|,H)} \right\} \right.$$

$$\left. + \mathbb{P}\left\{ \sum_r V_{X_i,r} < \frac{\gamma}{gPL(\|X_i\|,H)} \right\} \right)$$

$$= \mathbb{E}_\Phi \prod_{X_i \in \Phi} \left( \sum_{l_i \in \{L,N\}} p_{l_i}(y, H) \left( e^{-t} \mathbb{P}\left\{ \sum_r V_{X_i,r}^{l_i} \geq \frac{\gamma}{gPL(\|X_i\|,H)} \right\} + \mathbb{P}\left\{ \sum_r V_{X_i,r}^{l_i} < \frac{\gamma}{gPL(\|X_i\|,H)} \right\} \right) \right)$$

$$= e^{-2\pi\lambda(1-e^{-t}) \sum_{l \in \{L,N\}} \int_Z^\infty y p_l(y,H) \mathbb{P}\left\{\sum_r V_{y,r}^l \geq \frac{\gamma}{gPL_l(y,H)}\right\} dy}, \quad (5)$$

where in the second step, the LOS/NLOS status of communication links are independently drawn, and the last step is due to the Laplace generation functional (LGFL) of PPP. Using (5), the signal count is confirmed to be a Poisson random variable with the mean value:

$$\overline{\Sigma} = 2\pi\lambda \sum_{l \in \{L,N\}} \int_Z^\infty y p_l(y, H) \overline{F}_{\sum_n V_n^l}\left(\frac{\gamma}{gPL_l(y,H)}\right) dy. \quad (6)$$

Obtaining $\overline{\Sigma}$ in (6) requires the complementary cumulative distributed function (CCDF) of random variable, $\sum_r V_r^l$, namely $\overline{F}_{\sum_r V_r^l}(.)$. We then assume $\sum_r V_r^l \geq N \min_r V_r^l$ and further note that the fading power gains across antennas are i.i.d. normalized Gamma random variables, and therefore

$$\overline{F}_{\sum_n V_n^l}\left(\frac{\gamma}{gPL_l(y,H)}\right) \geq \left(\overline{F}_{V_1^l}\left(\frac{\gamma}{NgPL_l(y,H)}\right)\right)^N$$

$$= e^{-\frac{\gamma \overline{V}_l}{gPL_l(y,H)}} \left(1 - \sum_{m=0}^{\overline{V}_l} \frac{1}{m!} \frac{\overline{V}_l^m \gamma^m}{(NgPL_l(y,H))^m}\right)^N. \quad (7)$$

An upper-bound on $\overline{F}_{\sum_n V_n^l}(.)$ is then obtained using (6), (7).

We now use the approximated distribution of $\Sigma$ to study the (cumulative) pmf, i.e., $F_\Sigma(v) = \sum_{s=0}^v \frac{\overline{\Sigma}^s}{s!} e^{-\overline{\Sigma}}$. This quantity measures the probability that the signal count is smaller than a given number, and is plotted in Fig. 1 for high-rise and sub-urban environments. For a given receiver, it is highly likely to receive a large number of interfering signals. So, even where the antenna side-lobe gain is relatively very small, $G/g \gg 1$, since an exclusion zone fails to protect the ground receiver against the interference. Fig. 1 further shows that increasing the altitude of the outside-UAVs makes two completely different impacts on the pmf in the two communication environments: increasing $H$ reduces (increases) pmf in the sub-urban (high-rise) environment, i.e., $H$ increases due to multi-path dominance in the high-rise environment.

In practice, if $\Sigma$ becomes greater than a given threshold number, the ground receiver might report to the network. To preserve the receive quality, the network may adopt a combination of actions to mitigate/reduce the received interference. Here we assume that there is no such mechanism available, and instead attempt to gain quantitative insight on the design of the required size of exclusion zone, $Z^*$, that guarantees all outside-UAVs to be invisible to the receiver.

Let $O_Z$ denote the probability of $\Sigma > 0$. Our objective is then to evaluate $Z^*$, such that $O_Z \geq \epsilon$, where $\epsilon \in (0,1)$ is a

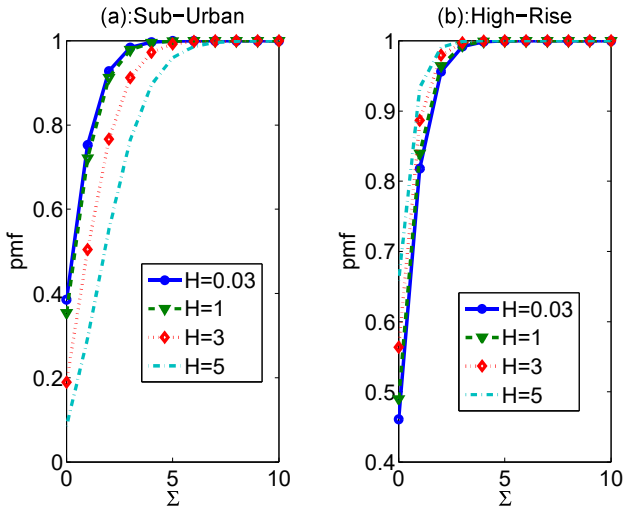

Fig. 1. The probability mass function (pmf) of the interference signal count, $\Sigma$, for $\lambda = 10^{-3}$, $G/g = 2500$, and $\gamma = 10^{-8}$.

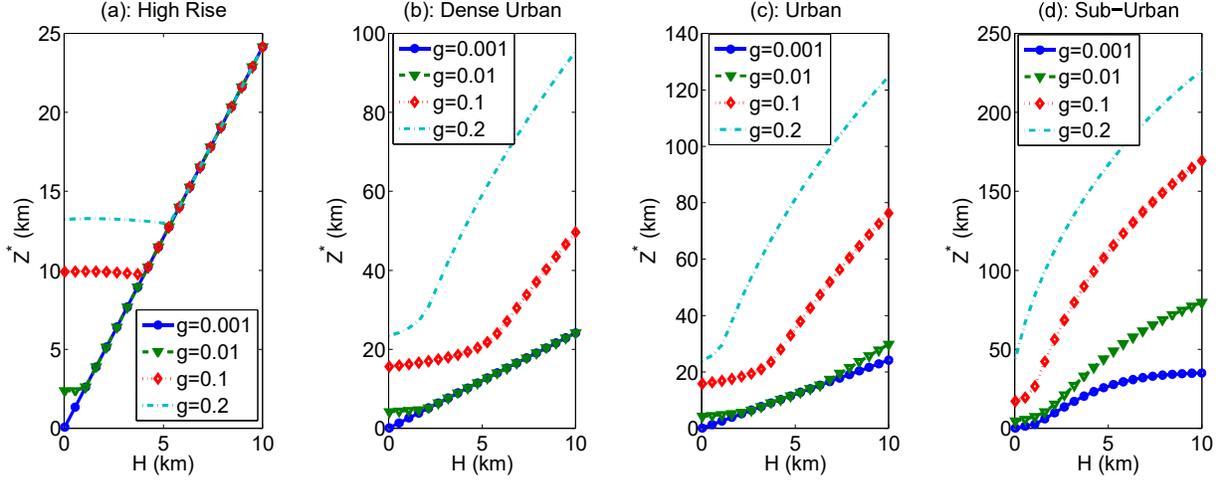

Fig. 2. Value of $Z^*$ versus the altitude of UAVs, where $\lambda = 10^{-3}$, and $\epsilon = 0.05$.

system parameter. Using (7), we write

$$O_Z \leq \exp\left\{ - 2\pi\lambda \sum_{l \in \{L,N\}} \int_Z^\infty y p_l(y,H) e^{-\frac{\gamma \overline{V}_l}{gPL_l(y,H)}} \right.$$
$$\left. \left(1 - \sum_{m=0}^{\overline{V}_l} \frac{1}{m!} \frac{\overline{V}_l^m \gamma^m}{(NgPL_l(y,H))^m}\right)^N dy \right\}.$$

Setting the upper-bound equal to $\epsilon$, $Z^*$ is obtained via the following equation

$$\sum_{l \in \{L,N\}} \int_{Z^*}^\infty \frac{y p_l(y,H)}{e^{\frac{\gamma \overline{V}_l}{gPL_l(y,H)}}} \left(1 - \sum_{m=0}^{\overline{V}_l} \frac{1}{m!} \frac{\overline{V}_l^m \gamma^m}{(NgPL_l(y,H))^m}\right)^N dy$$
$$= \frac{\log(1/\epsilon)}{2\pi\lambda}.$$

Fig. 2 plots $Z^*$ versus the altitude of interfering UAVs, $H$, in prevalent communication environments for several values of antenna side-lobe gain, $g$, where $\epsilon = 0.05$. Increasing the side-lobe gain, $g$, substantially increases $Z^*$. Fig. 2 also shows that even for a very small $g$, a rather large exclusion zone is required. The size of exclusion zone is also increased by increasing the UAVs' altitude.

Among the various type of communication environments, unlike the high-rise case, the sub-urban environment requires a substantially large exclusion zone, almost 4x larger as Fig. 2 shows, due mainly to the dominance of LOS component in the sub-urban environment.

## IV. IMPACT ON THE CAPACITY

Setting the size of exclusion zone based on the results in Section II reduces the negative impact of non-ideal antennas, i.e., $g \neq 0$, and limits the likelihood of introducing interfering signals at the receiver.

The performance of a multi-antenna receiver can also degrade due to the cross-antenna signal correlation. For the ideal case of $g = 0$, no cross-antenna signal correlation exists as the fading across antennas are independent. This enables the receive array to harness the diversity of the wireless channels and achieve diversity and/or multiplexing gains. In practice, where $g \neq 0$, the *aggregated interference* caused by the side-lobe induce cross-antenna signal correlation even when the exclusion zone $B_O(Z^*)$ is considered. In what follows, we investigate the capacity loss at the receiver because of the existing correlation in the received signal across the receive antennas.

### A. Cross-Antenna Interference Correlation

The cross-antenna interference correlation coefficient is:

$$\rho_{r,r'} = \frac{\mathbb{E}[I_r I_{r'}] - \mathbb{E}[I_r]\mathbb{E}[I_{r'}]}{\sqrt{\text{Var}(I_r)\text{Var}(I_{r'})}} = \frac{\mathbb{E}[I_r I_{r'}] - (\mathbb{E}[I_{r'}])^2}{\text{Var}(I_r)}, \quad (8)$$

where the received interference at antenna $r$ is

$$I_r = \sum_{X_i \in \Phi} PL(\|X_i\|, H) g V_{X_i, r}. \quad (9)$$

*Proposition 1:* The cross-antenna interference correlation coefficient is

$$\rho_{r,r'} = \frac{\sum_{l \in \{L,N\}} W_l}{\sum_{l \in \{L,N\}} \frac{\overline{V}_l + 1}{\overline{V}_l} W_l} = \frac{1}{1 + \frac{\sum_{l \in \{L,N\}} \frac{W_l}{\overline{V}_l}}{\sum_{l \in \{L,N\}} W_l}}, \quad (10)$$

where $r \neq r'$, and

$$W_l = \int_{Z^*}^\infty x p_l(x, H) (L_l(x, H))^2 dx. \quad (11)$$

*Proof:* See the Appendix. ∎

For any positive real numbers, $\{a_m, b_m\}$, $m = 1, 2, \ldots$, it is easy to show that $\sum_m \frac{a_m}{b_m} \geq \frac{\sum_m a_m}{\sum_m b_m}$. Therefore, $\frac{\sum_{l \in \{L,N\}} \frac{W_l}{\overline{V}_l}}{\sum_{l \in \{L,N\}} W_l} \leq$ $\sum_{l\in\{L,N\}} \frac{1}{\overline{V}_l}$. This suggests a lower-bound on $\rho_{r,r'}$ as $\rho_{r,r'} \geq \frac{1}{1+\sum_{l\in\{L,N\}}\frac{1}{\overline{V}_l}}$. An upper-bound on $\rho_{r,r'}$ is also obtained as

$$\rho_{r,r'} \leq \sum_{l\in\{L,N\}} \frac{1 W_l}{\frac{\overline{V}_l+1}{\overline{V}_l}W_l} = \sum_{l\in\{L,N\}} \frac{\overline{V}_l}{\overline{V}_l+1}.$$

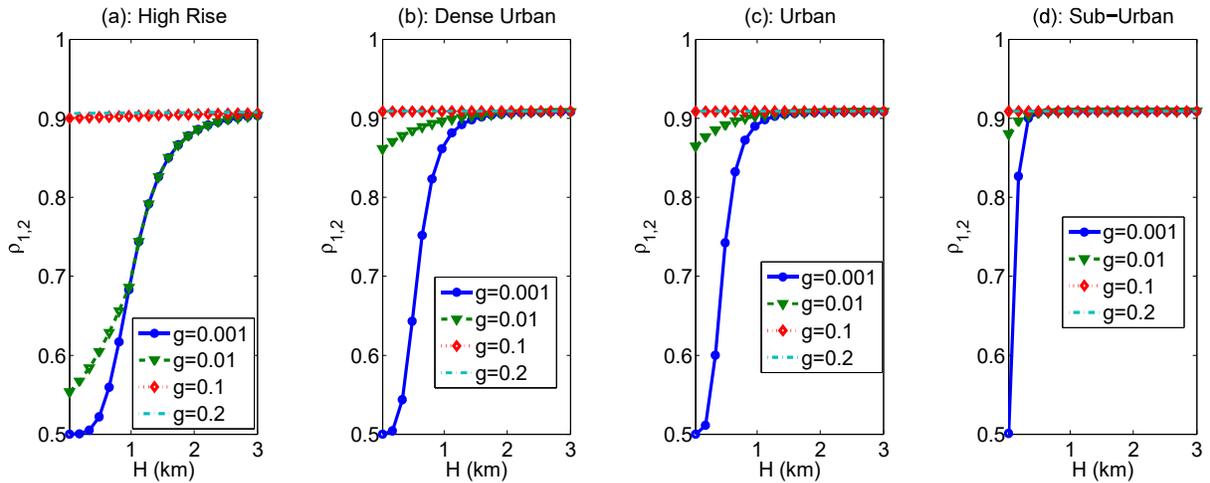

Fig. 3. Cross-antenna interference correlation coefficient versus the altitude of UAVs, where $\lambda = 10^{-3}$.

Therefore,

$$\frac{1}{1+\sum_{l\in\{L,N\}}\frac{1}{\overline{V}_l}} \leq \rho_{r,r'} \leq \sum_{l\in\{L,N\}} \frac{1}{1+\frac{1}{\overline{V}_l}}.$$

Using the above, for a Rayleigh distributed fading in the NLOS link, it is easy to show $\rho_{r,r'} \gtrapprox 0.5$.

Fig. 3 plots $\rho_{1,2}$ versus $H$, showing a considerable level of interference correlation, $\rho_{1,2} \geq 0.5$, despite consideration of exclusion zone to protect the receiver from the direct interference received from the outside-UAVs. Such a correlation limits the achievable gain of the receive diversity in the A2G communication link. Increasing $g$ also leads to a substantial increase of $\rho_{1,2}$. Fig. 3 further highlights the impact of the communication environment. A much higher antenna correlation is experienced in the sub-urban environment due to the dominance of the LOS interference.

### B. Capacity Loss

The supporting UAV, denoted by $X_0$, is located at height $\tilde{H}$ on top of the receiver. We assume that the same channel model discussed in Section II is valid for the communication link between the supporting UAV and the ground receiver. The achievable rate of the receiver is denoted by $C(g)$.

We evaluate $\Delta_R = C(0) - C(g)$ for a given system with an exclusion zone of $B_O(Z^*)$. Assuming maximal ratio combining, which is optimal for the case of no interference [16], $\Delta_R$ is then formulated as

$$\Delta_R = \mathbb{E}\left[\log\left(1+\sum_{r=1}^{R}\frac{PL(\|X_0\|,\tilde{H})GV_{X_0,r}}{\sigma^2}\right) - \log\left(1+\sum_{r=1}^{R}\frac{PL(\|X_0\|,\tilde{H})GV_{X_0,r}}{\sigma^2+I_r}\right)\right]$$

$$\approx \mathbb{E}\left[\log\left(1+\frac{PL(\|X_0\|,\tilde{H})G\sum_{r=1}^{R}V_{X_0,r}}{\sigma^2}\right) - \log\left(1+\frac{PL(\|X_0\|,\tilde{H})G\sum_{r=1}^{R}V_{X_0,r}}{\sigma^2+\hat{I}}\right)\right],$$

where $\sigma^2$ is the AWGN power, the interference received at each antenna is approximated as $\hat{I} = \sum_{X_i\in\Phi}PL(\|X_i\|,H)gV_{X_i}$, and $V_{X_i}$ is distributed as in (3) and does not depend on the antenna index. This is due to highly correlated interference across antennas, as substantiated in Proposition 1.

Using $\log(1+x) = \int_0^{\infty} e^{-v}/v\left(1-e^{-vx}\right)dv$, we then write $\Delta_R$ as:

$$\Delta_R \approx \int_0^{\infty} e^{-v\sigma^2}/v\left(1-\mathcal{L}_{\hat{I}}(v)\right)(1-\mathcal{L}_S(v))dv. \quad (12)$$

where $\mathcal{L}_{\hat{I}}(v)$ is the Laplace transform of $\hat{I}$,

$$\mathcal{L}_{\hat{I}}(v) = \mathbb{E}[e^{-v\hat{I}}] = \mathbb{E}_\Phi \prod_{X_i\in\Phi} \mathbb{E}e^{-vgPL(\|X_i\|,H)V_{X_i}}$$

$$= \mathbb{E}_\Phi \prod_{X_i\in\Phi}\left(\sum_{l_i\in\{L,N\}}\frac{p_{l_i}(\|X_i\|,H)}{(1+vPgL_{l_i}(\|X_i\|,H)/\overline{V}_{l_i})^{\overline{V}_{l_i}}}\right)$$

$$= e^{-2\pi\lambda\sum_{l\in\{L,N\}}\int_{Z^*}^{\infty} yp_l(y,H)\left(1-(1+vPgL_l(y,H)/\overline{V}_{l_i})^{-\overline{V}_l}\right)dy}. \quad (13)$$

In (12), $\mathcal{L}_S(v)$ is the Laplace transform of the effective received power from the supporting UAV $X_0$:

$$\mathcal{L}_S(v) = \sum_{l \in \{L,N\}} \frac{p_l(0, \tilde{H})}{(1 + vGPK_l\tilde{H}^{-\alpha_l})^{\overline{V}_l}}.$$

Fig. 4 plots $\Delta_R$ versus $H$ for several values of $g$ in different communication environments, showing that the loss of capacity is very small. In effect, the inclusion of the exclusion zone is effective in minimizing the impact of interference, which is consistent across all communication environments and values of $g$. We further see that in general increasing $H$ can increase capacity loss or decrease capacity loss, depending on communication environment and $g$. In fact, a combination of factors including the likelihood of receiving LOS interference, higher path-loss attenuation, and the size of exclusion zone affect the capacity loss. Depending on this combination, the capacity loss may increase or decrease by increasing $H$. Nevertheless, for all cases the capacity loss always stays restricted. This is mainly because the chance that a dominant LOS interference exists in the exclusion zone vanishes. Besides, the inclusion of well-designed exclusion zone weakens the effect of NLOS interfering links.

## V. Conclusions

We have investigated the impact of UAVs' antenna sidelobe on the performance of UAV-enabled communication. Our analysis and results demonstrated that even for a very small value of the antenna's side-lobe gain, the ground receiver can experience substantial interference. We further showed that a rather large exclusion zone is required to ensure a sufficient level of protection for the ground receiver. Nevertheless, in a multiple-antenna setting for the ground users, even when such large exclusion zone was in place, UAVs' antenna sidelobe creates a high level of correlation among the interference signals received across receive antennas. Such a correlation among the received signals limits the system's ability to exploit channel diversity in a multiple-antenna setting for improving capacity. Using these results/observations, we then quantified the impact of UAVs' antenna side-lobes on the overall system performance by obtaining the corresponding loss on the achieved capacity in various communications environments. We observed that the capacity loss can be limited by careful selection of system parameters. Finally, our results indicate that unlike the terrestrial communications in which non-LOS propagation is often dominant, designing UAV communication systems without considering a realistic antenna pattern might lead to substantial capacity loss in real settings.

## Appendix: Proof of Proposition 1

Using Campbell-Mecke Theorem [17],

$$\mathbb{E}[I_r] = g\mathbb{E} \sum_{X_i \in \Phi} PL(\|X_i\|, H)V_{X_i, r}$$

$$= 2\pi P g \lambda \int_{Z^*}^{\infty} x\mathbb{E}[L(x,H)V_{x,r}]dx$$

$$= 2\pi g P \lambda \sum_{l \in \{L,N\}} \int_{Z^*}^{\infty} xp_l(x,H)L_l(x)dx, \quad (14)$$

where we note $\mathbb{E}[V_l] = 1$. Further, for $r \neq r'$

$$\mathbb{E}[I_r I_{r'}] = g^2 P^2 \mathbb{E} \sum_{X_i \in \Phi} \sum_{X_j \in \Phi} L(\|X_i\|, H)$$

$$\times L(\|X_j\|, H)V_{X_i, r}V_{X_j, r'}$$

$$= g^2 P^2 \mathbb{E} \sum_{X_i \in \Phi} (L(\|X_i\|, H))^2 V_{X_i, r}V_{X_i, r'}$$

$$+ g^2 P^2 \mathbb{E} \sum_{X_i \in \Phi} \sum_{X_j \in \Phi \setminus X_i} L(\|X_i\|, H)L(\|X_j\|, H)V_{X_i, r}V_{X_j, r'}$$

$$= 2\pi g^2 P^2 \lambda \sum_{l \in \{L,N\}} \overline{V}_l^2 \int_{Z^*}^{\infty} xp_l(x,H)(L_l(x,H))^2 dx$$

$$+ 4\pi^2 g^2 P^2 \lambda^2 \sum_{l \in \{L,N\}} \sum_{l' \in \{L,N\}} \overline{V}_l \overline{V}_{l'} \int_{Z^*}^{\infty} \int_{Z^*}^{\infty} x_1 x_2$$

$$p_l(x_1, H)L_l(x_1, H)p_{l'}(x_2, H)L_{l'}(x_2, H)dx_1 dx_2$$

$$= 2\pi g^2 P^2 \lambda \sum_{l \in \{L,N\}} \int_{Z^*}^{\infty} xp_l(x,H)(L_l(x,H))^2 dx$$

$$+ 4\pi^2 g^2 P^2 \lambda^2 \sum_{l \in \{L,N\}} \sum_{l' \in \{L,N\}} \int_{Z^*}^{\infty} \int_{Z^*}^{\infty} x_1 x_2$$

$$p_l(x_1, H)L_l(x_1, H)p_{l'}(x_2, H)L_{l'}(x_2, H)dx_1 dx_2. \quad (15)$$

Using (15), and (14), it is easy to show

$$\mathbb{E}[I_r I_{r'}] - (\mathbb{E}[I_{r'}])^2 =$$

$$2\pi g^2 P^2 \lambda \sum_{l \in \{L,N\}} \int_{Z^*}^{\infty} xp_l(x,H)(L_l(x,H))^2 dx. \quad (16)$$

Similarly, noting that $\mathbb{E}[(V_l)^2] = \frac{\overline{V}_l + 1}{\overline{V}_l}$,

$$\text{Var}(I_r) = 2\pi g^2 P^2 \lambda \sum_{l \in \{L,N\}} \frac{\overline{V}_l + 1}{\overline{V}_l} W_l. \quad (17)$$

Substituting (16) and (17) into the definition of cross-antenna correlation in (8) completes the proof. ∎

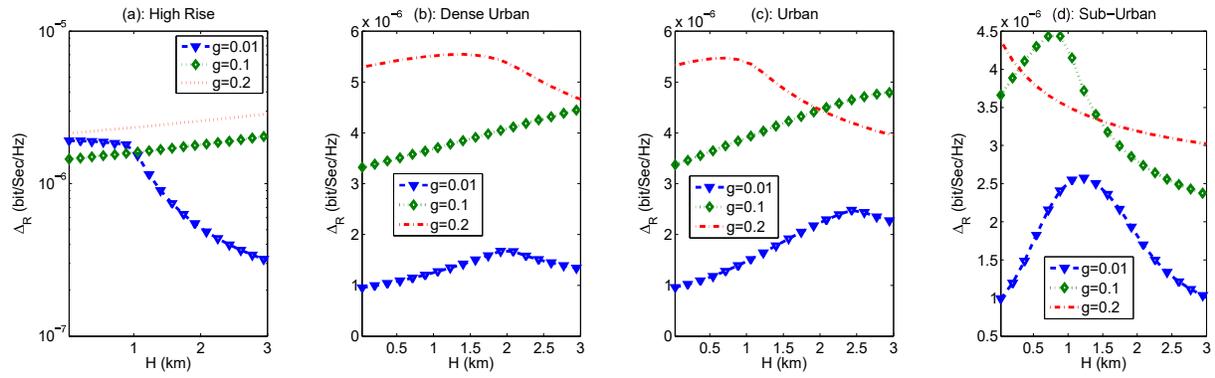

Fig. 4. Capacity loss versus $H$, where $\tilde{H} = 300$ m, and $\lambda = 10^{-3}$.